\documentstyle[preprint,aps,prl]{revtex}
\begin{document}
\draft
\title{Many-body and Covalence Effects in the Polarization \\ of Ferroelectric
Perovskites}

\author{R. Resta$^{(1,2)}$ and S. Sorella$^{(2)}$}
\address{$^{(1)}$ Dipartimento di Fisica Teorica, Universit\`a di
Trieste, Strada Costiera 11,  I-34014 Trieste, Italy \\
$^{(2)}$ Scuola Internazionale Superiore di Stud\^\i\
Avanzati (SISSA), Via Beirut 4, I-34014 Trieste, Italy}

\maketitle

\begin{abstract} The ferroelectric polarization of perovskite oxides is much
larger than implied by displacement of static ionic charges. We use an
explicitly correlated scheme to investigate the phenomenon; charge transport is
evaluated as a geometric quantum phase. Both covalence and electron-electron
interaction enhance polarization in the weakly correlated regime. At higher
values of the electron-electron interaction, the system undergoes a transition
from a band insulator to a Mott insulator: the static ionic charge is
continuous across the transition, whereas the polarization is discontinuous.
Above the transition, oxygen transports a {\it positive} charge. \end{abstract}

\pacs{~ \\ SISSA Preprint \# 174/94/CM/MB}

\narrowtext
\newpage

The large values of the spontaneous polarization in ferroelectric perovskites
cannot be explained on the basis of a simple ionic model. For instance, the
effective dynamical charge associated to an oxygen displacement during the
ferroelectric distortion is of the order of $-$6, {\it i.e.} about three times
the nominal static value given by a completely ionic picture
\cite{rap75,King94b} (atomic units of charge are adopted throughout).
First-principles calculations, performed within the local-density approximation
(LDA), have demonstrated that this effect is due to a large amount of covalence
\cite{rap84}: as a basic feature of such calculations, the effect of
electron-electron interaction is accounted for in a mean-field way. In the
present paper we investigate a different facet of the problem, namely how the
picture is changed by explicitly considering correlated wavefunctions: such
investigation is based on a completely different theoretical and computational
scheme. In the moderately correlated regime, we find that the electron-electron
interaction reduces the static ionic charges but---in apparent
contradiction---significantly enhances the macroscopic polarization. At higher
values of the electron-electron interaction the system has a transition from a
band insulator to a Mott insulator, first investigated by Egami {\it et al.}
\cite{Egami}. When we examine the transition from the standpoint of macroscopic
polarization, several exotic phenomena show up. In particular the static ionic
charges are continuous, whereas the polarization is discontinuous: therefore
the polarization (or equivalently the associated geometric quantum phase
\cite{Rapix,Ortiz94}) is the primary order parameter for the transition.

The ferroelectric perovskites have formula unit ABO$_3$, where A is a mono- or
divalent cation, and B is a penta- or tetravalent cation in the transition
series. The simplest ferroelectric structure is the tetragonal one, where the
most important symmetry-breaking distortion is a relative displacement of the B
sublattice with respect to the O one. The essence of the phenomenon can be
schematized by considering only the linear O--B chain along the ferroelectric
axis, which is symmetric in the paraelectric (so-called prototype) structure,
and asymmetric in the ferroelectric one. In order to make a correlated
wavefunction available, we schematize the Hamiltonian of such a linear model
ferroelectric with a  two-band Hubbard model at half filling, first introduced
by Egami et {\it al.} \cite{Egami}, and whose macroscopic polarization has not
been investigated so far. In order to reduce the number of parameters, we
arbitrarily set equal values of the Hubbard $U$ on anion and cation. The
Hamiltonian of the prototype structure is therefore \begin{equation} \sum_{j
\sigma} [ \; (-1)^j \Delta \, c^\dagger_{j \sigma} c_{j \sigma} - t_0 (
c^\dagger_{j \sigma} c_{j+1 \sigma} + \mbox{\rm H.c.} ) \; ] + U \sum_j
n_{j\uparrow}  n_{j\downarrow} \; , \end{equation} and depends on two
parameters besides $U$: the hopping $t_0$, and  the difference in site energies
$E_{\rm B} - E_{\rm O} = 2 \Delta $. We restore charge neutrality by assigning
a classical charge of +2 to the cationic sites: this ensures that in the
extreme ionic limit ($t_0 = 0$ and $U < 2 \Delta$) the total static charges are
$\pm 2$. The energy difference  $E_{\rm B} - E_{\rm O}$ is of the order of 3 or
4 eV in the materials of interest \cite{HarrisonNew}, taking {\it e.g.} Ti or
Nb as the B cation; it also equals the gap of the noninteracting model.  In the
following of this study we assume $\Delta = 2$ eV, and $t_0 = 3.5$ eV, such as
to get a mean-field band width of 5.3 eV, which is a realistic value for the
valence bands of these materials. The fact that $t_0$ is of the order of
2$\Delta$ clearly indicates a mixed ionic/covalent character.

We plot in Fig.~\ref{f1}(a) the static ionic charge as a function of U
(triangles). The calculations have been performed in 8-sites supercells, using
skew (quasiperiodic) boundary conditions on each electronic variable, and then
taking the average over the boundary conditions (alias over the supercell
quasimomenta $k$), as in Ref.~\cite{Gros92}. This ensures that the $U
\rightarrow 0$ limit is numerically equal to the fully converged thermodynamic
limit of the noninteracting calculation.  Typically we have used 30 $k$ points,
corresponding to 120 $k$ points in the unfolded Brillouin zone of the
noninteracting system. The numerical diagonalization was performed via the well
known Lanczos algorithm, which provided the ground state electronic
wavefunctions $\Psi_0(k)$  with an energy tolerance close to the machine
accuracy ($10^{-12} t_0$ ), with less than $100$  Lanczos iterations in {\em
all} cases.  In order to reduce the problem size we have explicitly used the
conservation of the  number of spin-up $N_\uparrow$ and spin-down
$N_\downarrow$ particles.  The subspace with $N_\downarrow=N_\uparrow=4$, where
the ground state lies, contains  only  $4900$ elements, thus allowing  a  large
reduction of  the full Hilbert space (amounting to $4^8=65536$ elements).   Use
of the center-of-mass translation symmetry is also possible, but becomes useful
only for larger systems. In an insulating system as the present one the size
effects are  small, and further minimized by $k$-averaging \cite{Gros92}. For a
few parameter values we have indeed performed 12-sites calculations, and
checked
that the results agree with the 8-sites ones to within one per cent, at least
for the quantities studied here.

The ferroelectric distortion is the zone-center optical phonon, where the site
coordinates are displaced by a relative amount $\xi$, typically of the order of
$\xi_{\rm F} = 0.05 a$, where $a$ is the lattice constant: this distortion
asymmetrically affects the hopping matrix elements. We assume the other
parameters fixed during the ferroelectric distortion, while for the $t$
variation we assume the simple Su-Schrieffer-Heeger \cite{Su} linear dependence
$t = t_0 \pm 2 \alpha \xi$. For reasons given below, a realistic modeling of
these materials requires a rather large electron-phonon coupling $\alpha$,
typically $\alpha a \simeq 10$ eV in the mean-field case. The  tight-binding
noninteracting Hamiltonian is trivially diagonalized as \begin{equation}
\epsilon(k) = \pm \sqrt{\Delta^2 + 4 t_0^2 \cos^2{ka/2} + 16 (\alpha \xi)^2
\sin^2{ka/2}} \; . \end{equation} The band structure is quadratic in $\xi$,
hence the (linear) deformation potential vanishes in the prototype structure.
Nonetheless the band shift induced by the actual ferroelectric distortion
$\xi_{\rm F}$ is rather large (about $-$0.8 eV at the zone boundary) in
semiquantitative agreement with LDA calculations (see {\it e.g.}
Fig.~\ref{f1}(a) in Ref.~\cite{rap84}).

The static charges are somewhat reduced by the distortion, as shown in
Fig.~\ref{f1}(a), circles. When $U$ is increased to large values, the system
undergoes an interesting transition, from a band insulator to a Mott insulator.
With our parameter values, the transition occurs at $U_{\rm C} = 2.27 t_0$.
Egami {\it et al.}, who first investigated this transition using only $k = 0$
wavefunctions, found a discontinuous drop in the static ionic charges of the
prototype structure, while the charges of the ferroelectric structure were
found continuous as a function of $U$. We reproduce their results, but we also
find that that the discontinuity disappears as $k$-point convergence is
approached. A careful analysis is displayed in Fig.~\ref{f1}(b), which
incidentally proves the effectiveness of the $k$ average \cite{Gros92} in
getting rid of spurious finite-size effects. The apparent discontinuity is due
to a level crossing which occurs at $k = 0$ and {\it not} at $k \neq 0$, as
discussed below. We have explicitly verified that the computed discontinuity is
inversely proportional to the number of $k$ points used. Furthermore the
discontinuity disappears even with a coarse mesh if the mesh is displaced on
the $k$ axis in order to avoid the $k = 0$ singular point: this is also shown
in  Fig.~\ref{f1}(b), open circles.

We are interested in the macroscopic polarization $\Delta P$ induced by the
distortion, when the sites are continuously displaced from the prototype
structure ($\xi = 0$) to the ferroelectric one ($\xi = \xi_{\rm F}$). We
therefore need evaluating how much charge is transported along the chain during
a relative displacement of the two sublattices, in a vanishing electric field
\cite{Rapix}; if we choose to keep the origin fixed on a cationic site, the
transport is purely electronic. The electronic charge transport is best
evaluated as a geometric quantum phase, as first  shown---for the explicitly
correlated case---by Ort\'{\i}z and Martin \cite{Ortiz94}. The rationale behind
the geometric phase approach is that the dynamical charge is a
quantum-mechanical current---hence a phase of the wavefunction---and bears in
general {\it no relationship} to the modulus of the wavefunction when periodic
boundary conditions are used.

The starting ingredients are again the same ground-state wavefunctions
$\Psi_0(k)$ discussed above. One then removes the Bloch-like phase upon
defining $\Phi_0(k) = \Psi_0(k) \prod \exp (- i k x_j)$, where $x_j$ are the
site coordinates. The wavefunctions $\Phi_0(k)$ so obtained are {\it periodic}
over the supercell at any $k$, and implicitly depend on the Hamiltonian
parameters. Only the dependence upon $\xi$ and $U$ will be relevant in the
following discusssion. One then defines the many-body generalization of the
geometric phase first introduced by Zak \cite{Zak}: \begin{equation}
\gamma(\xi,U) = i \int dk \; \langle \Phi_0(k) | \frac{\partial}{\partial k}
\Phi_0(k) \rangle, \end{equation}  where the $k$-integration is over the
Brillouin zone of the supercell. The numerical calculation proceeds as in the
uncorrelated case \cite{Rapix}, and the macroscopic polarization of the
ferroelectric structure is \begin{equation} \Delta P = [ \gamma(\xi_{\rm F},U)
- \gamma(0,U) ] / 2 \pi , \label{pol} \end{equation} defined modulo a
polarization quantum of magnitude 1, corresponding to the transport of one
charge over one cell. This is one half of the quantum of the mean-field theory
\cite{Rapix}, where double occupancy of one-particle states is enforced.

In the special case where the electron-phonon coupling $\alpha$ is taken as
vanishing, then each site may only transport its static ionic charge (shown in
Fig.~\ref{f1}): in fact the polarization  calculated as a geometric phase
accounts
precisely for this rigid charge transport. Notice however that the two
alternative calculations are not {\it numerically} equivalent, thus providing a
useful convergence test. Using our typical $k$-point mesh given above, the
error is smaller than 10$^{-3}$. When $\alpha \neq 0$, the dynamical charge is
no longer equal to the static one, and is typically much larger than it.

Let us first illustrate the mean-field ($U = 0$) calculation. We take $\alpha a
= 10$ eV, which provides a spontaneous polarization in agreement with both
measurements and LDA calculations in typical ferroelectric perovskites. The
polarization $\Delta P$ is almost linear in $\xi$: the relevant quantities to
display are therefore  the average dynamical charge $\langle Z^*(\xi) \rangle =
a \Delta P(\xi) / \xi $ and the linear (or Born) dynamical charge of the
prototype structure $Z^* = a P'(0) $. The static (cationic) charge is 1.47:
charge transport is enhanced by a factor larger than four by the
electron-phonon coupling, thus providing giant dynamical charges, and large
values of the spontaneous polarization. The actual values within our model are
$\langle Z^*(\xi_{\rm F}) \rangle = 5.95$ and $Z^* = 7.28$.

We now switch discussing the effects of electron-electron interaction. The
first interesting phenomenon occurs already in the centrosymmetric system,
having real wavefunctions $\Phi_0(k)$: the Zak phase $\gamma$ changes
discontinuously by $\pi$ at the transition point. Equivalently, one finds a
Berry phase of $\pi$ around the rectangular loop in the $(k,U)$ plane shown in
Fig.~\ref{f3}, since the vertical sides of the rectangle do not contribute
\cite{Rapix,Ortiz94}. This means that the real wavefunction undergoes a sign
change when transported along the closed path: the commonest occurrence of such
a feature, well known in molecular physics \cite{Mead92}, is due to the
presence of a point of double degeneracy inside the domain. This is precisely
the case here: there are two well distinct states whose energies cross at the
point $(0,U_{\rm C})$, whereas at $k \neq 0$ there is no degeneracy. We have
numerically checked the level crossings by exploiting the metastability of the
Lanczos iteration across the transition. Coming now to the physical meaning of
such transition, we notice that a phase change of $\pi$ corresponds to the
transport of an electronic charge over half a lattice constant, from an oxygen
site to a cationic one. Notice once more that such transport occurs without
affecting the static charges. This is a virtue of the ring geometry of our
chain, whereas in a finite linear chain, owing to continuity, charge transport
would obviously affect the static charges of the end sites. We thus discover
that the geometric phase---and not the static ionic charge---is the primary
order parameter for the transition. One could even straightforwardly generalize
the ``band-center operator'' of Ref.~\cite{Zak} to the many-body case, and
characterize the transition by saying that the crystalline ground state is an
eigenstate of such operator belonging to different eigenvalues below and above
$U_{\rm C}$.

We then consider the polarization of the ferroelectric structure, calculated as
in Eq.~(\ref{pol}) and where the  $\Phi_0(k)$ are complex. We plot in
Fig.~\ref{f2} the average dynamical charge  $\langle Z^* \rangle$ as a function
of $U$, for several values of $\xi$. In the moderate-$U$ region below $U_{\rm
C}$ the electron-electron interaction produces a significant enhancement of the
polarization. Notice that this latter feature is in apparent contradiction with
the fact that the static ionic charges {\it decrease} instead with increasing
$U$. The most prominent feature of Fig.~\ref{f2} is the divergence at $U_{\rm
C}$, which has an interesting physical meaning. The (near) divergent curve,
corresponding to our smallest $\xi$, is an approximation to the Born dynamical
charge: the figure then indicates that at $U = U_{\rm C}$ an infinitesimal
sublattice displacement, (starting from the prototype structure) induces a {\it
finite} charge transport, hence an infinite $Z^*$. At finite $\xi$ values
instead---and in particular at the value $\xi_{\rm F} = 0.05a$ corresponding to
a realistic ferroelectric distortion---the polarization has a finite and large
discontinuity at $U_{\rm C}$. Notice that  the Zak phase of the ferroelectric
structure is continuous as a function of $U$, and therefore both the divergence
and the discontinuity  of the dynamical charges must be traced back to the
discontinuity of the centrosymmetric ($\xi = 0$) Zak phase in Eq.~(\ref{pol}).

In conclusion, we have investigated here the effect of electron-electron
interaction in the macroscopic polarization of ferroelectric perovskites, by
means of an explicit model Hamiltonian which captures the basic features of the
phenomenon, and exploiting the geometric phase approach. The very large
polarization of these materials owes to their mixed ionic/covalent character,
and is further enhanced by electron-electron interaction, as long as its
strength remains moderate. At higher strength, the system undergoes a
transition from a band insulator to a Mott insulator. At the transition point,
the polarization is discontinuous, and even reverses its sign for a given
sublattice displacement. In the highly correlated regime the cation transports
a negative dynamical charge, and oxygen a positive one.

{}~~

We are grateful to E. Tosatti for many helpful discussions and for a critical
reading of the manuscript. Part of this work was performed while the authors
were at the Institute for Theoretical Physics, University of California at
Santa Barbara. The research was supported in part by National Science
Foundation under Grant No. PHY89-04035.

\begin{figure} \caption{Static charge of the cation as a function of the
Hubbard $U$. (a) Prototype centrosymmetric structure (triangles) and
ferroelectric structure (circles) below the transition. (b) Different
computations for the prototype structure at the transition point $U_{\rm C}$.
Empty triangles: 10 $k$ points. Filled triangles: 100 $k$ points. Empty
circles: 30 $k$ points,  displaced such as to avoid $k = 0$.} \label{f1}
\end{figure}

\begin{figure} \caption{Average dynamical charge of the cation as a function
of the Hubbard $U$, for different values of the displacement $\xi$. In order of
increasing value of the discontinuity at $U_{\rm C}$, the curves represent:
$\xi = \xi_{\rm F} = 0.05 a$; $\xi = 0.035 a$;  $\xi = 0.0245 a$;  $\xi = 0.014
a$;  $\xi = 0.0035 a$. } \label{f2}  \end{figure}

\begin{figure} \caption{Rectangular loop in the $(k,U)$ plane which encircles
the level crossing at $k$=0 and $U$=$U_{\rm C}$, in arbitrary units. The
projection over the $k$ axis coincides with the Brillouin zone of the
supercell. The cross is at the degeneracy point.} \label{f3}  \end{figure}

\end{document}